\documentclass[preprint,12pt]{elsarticle}

\usepackage{graphicx}
\usepackage{amssymb}

\usepackage{lineno}

 \usepackage {paralist}
 \usepackage{subfig}
\usepackage{rotating, graphicx}
\usepackage{amsmath,amssymb,amsfonts}
\usepackage{algorithmic}
\usepackage{multirow}
\usepackage{textcomp}
\usepackage{xcolor}
\usepackage{booktabs,threeparttable}
\usepackage{array} 
\usepackage{url}

\usepackage{xcolor}
\usepackage{ifthen}
\newboolean{showcomments}
\setboolean{showcomments}{false} 
\ifthenelse{\boolean{showcomments}}
    {
        \newcommand{\thomas}[1]{\textcolor{teal}{{\it [Thomas says: #1]}}}
        \newcommand{\johan}[1]{\textcolor{red}{{\it [Johan says: #1]}}}
        \newcommand{\per}[1]{\textcolor{magenta}{{\it [Per says: #1]}}}
        \newcommand{\change}[1]{\textcolor{blue}{{#1}}}
    }
    {
        \newcommand{\thomas}[1]{}
        \newcommand{\johan}[1]{}
        \newcommand{\per}[1]{}
        \newcommand{\change}[1]{#1}
    }

\journal{Journal of Systems and Software}

\begin{document}

\begin{frontmatter}


\title{Open Data Ecosystems -- an empirical investigation into an emerging industry collaboration concept}



\author[lu]{Per Runeson $^{*,}$}
\ead{per.runeson@cs.lth.se}
\author[rise]{Thomas Olsson}
\ead{thomas.olsson@ri.se}
\author[lu]{Johan Linåker}
\ead{johan.linaker@cs.lth.se}

\address[lu]{Dept. of Computer Science, Lund University, Lund, Sweden}
\cortext[cor1]{Corresponding author}

\address[rise]{Systems Engineering, RISE Research Institutes of Sweden AB, Lund, Sweden}

\begin{abstract}
Software systems are increasingly depending on data, particularly with the rising use of machine learning, and developers are looking for new sources of data. Open Data Ecosystems (ODE) is an emerging concept for data sharing under public licenses in software ecosystems, similar to Open Source Software (OSS). It has certain similarities to Open Government Data (OGD), where public agencies share data for innovation and transparency. 

We aimed to explore open data ecosystems involving commercial actors. Thus, we organized five focus groups with 27 practitioners from 22 companies, public organizations, and research institutes. Based on the outcomes, we surveyed three cases of emerging ODE practice to further understand the concepts and to validate the initial findings. The main outcome is an initial conceptual model of ODEs' value, intrinsics, governance, and evolution, \change{and propositions for practice and further research}.

We found that ODE must be value driven. Regarding the intrinsics of data, we found their type, meta-data, and legal frameworks influential for their openness.  
We also found the characteristics of ecosystem initiation, organization, data acquisition and openness be differentiating, which we advise research and practice to take into consideration.  

\end{abstract}


\end{frontmatter}



\section{Introduction}
Open innovation and co-creation are ways for organizations to leverage the creativity outside the own organizational boundaries. Chesbrough coined the term Open Innovation (OI) in 2003, initially referring to exchange of ideas. OI is {``a paradigm that assumes that firms can and should use external ideas as well as internal ideas\dots as they look to advance their technology''}~\cite{chesbrough2003open}. Later, Chesbrough \emph{et al.~}redefined OI as {``a distributed innovation process\dots across organizational boundaries, using pecuniary and non-pecuniary mechanisms''}~\cite{chesbrough2014new}. Open innovation is manifested in software engineering through Open Source Software (OSS)~\cite{LinakerJSS2018} and software ecosystems~\cite{JANSEN20121495}.

Development and operation of software systems have become increasingly dependent on data during the last decade~\cite{gandomi2015beyond,CoyleValue2020}. In particular Machine Learning (ML) applications require lots of high-quality data, while traditional systems use data to provide services to its users. Raj \emph{et al.~}identify data management challenges, such as shortage of data, need for sharing techniques, and data quality~\cite{raj2019data}. As suggested in our previous work, co-creation and collaboration principles have to be adopted to harness the innovation potential and to manage costs in the age of data~\cite{Runeson19}. This is in line with other researchers' observations of needs for ecosystem strategies when working with open data~\cite{rudmark2019harnessing}.

Examples of such co-creation and collaboration can be found in the domain of OSS, which is utilized in almost all software systems, and is commonly integrated with commercial offerings. In software ecosystems~\cite{JANSEN20121495}, OSS is a means to share platform software and tools with partners -- and even competitors -- both to reduce cost and to promote OI. This involves trade-offs between what software to share and what to keep proprietary~\cite{LinakerEMSE2020}. Extending similar practices to data have so far primarily been initiated by public agencies. 
Open Government Data (OGD), i.e. public agencies giving access to public data, is brought forward as an enabler for innovation and entrepreneurship, both by politicians and researchers~\cite{lakomaa2013open,dawes2016planning}, and is studied quite intensively~\cite{attard2015systematic}. Recently, the Bennett Institute for Policy, Cambridge, launched a report on ``The Value of Data''\cite{CoyleValue2020} with a focus on public policy for data. They conclude that {``[v]alue comes from data being brought together, and that requires organisations to let others use the data they hold.''}  However, as far as we have seen, the opening of data between commercial organizations to create more value, with or without governmental involvement, is not practiced to any major extent, with some reported exceptions (see e.g., Susha \emph{et al.}~\cite{susha2020towards}).

We want to advance knowledge about data in conjunction with OI by exploring the concept of Open Data Ecosystems (ODE), and how they align with industrial and governmental practices. We, therefore, launched a focus group series on attitudes and expectations on collaborating with external organizations on data. Secondly, we conducted a case survey~\cite{PetersenCaseSurveyChapter2020} of three emerging ODEs, to study how the ODE concept manifests in practice. The initial results from the focus group study are presented in a conference paper~\cite{RunesonSEEA2020}. In this paper, we add the experiences from three emerging ODEs and integrate it with existing literature into a conceptual model for ODEs\footnote{We used the term Open Data Collaboration (ODC) in the conference paper~\cite{RunesonSEEA2020}, but as the concept of Data Ecosystems emerges in the literature~\cite{Oliveira19} we adapt to that terminology.}. 

The concept of Data Ecosystem is emerging in the literature~\cite{Oliveira19}, although not uniquely defined. We define the concept of data ecosystems inspired from software ecosystems~\cite{jansen2020focus} and open government data~\cite{zuiderwijk2014innovation, Oliveira19}: 
\textit{ 
\begin{itemize}
    \item[] A data ecosystem is
    \begin{itemize}
        \item[--] a networked \emph{community of actors} (organizations and individuals), which base their relations to each other on a \emph{common interest}~\cite{zuiderwijk2014innovation}, 
        \item[--] supported by an underpinning \emph{technological platform}~\cite{jansen2020focus}
        \item[--] that enables actors to process data (e.g., find, archive, publish, consume, or reuse) as well as to foster innovation, create value, or support new businesses~\cite{Oliveira19}. 
        \item[--] Actors \emph{collaborate on the data and boundary resources} (e.g., software and standards), through the exchange of information, resources, and artifacts~\cite{jansen2020focus}.
    \end{itemize}
\end{itemize}}

The Open Data Institute has defined a spectrum of openness for data ecosystems, from \emph{closed} to \emph{shared} to \emph{open}~\cite{CoyleValue2020}. We thus refer to ODE as data ecosystems including data across this spectrum. We are interested in data ecosystems with both commercial and public organizations as actors. Since our initial observations point to the central role of technological platforms we include them as well in the definition. 

There are several challenges with ODE -- legal, organizational, and technical. For example, adhering to privacy laws when data is shared across organizations, choosing business models and strategies for when to share data and when to keep it as a competitive advantage, and finding technical solutions for sharing data in secure and efficient ways. 

We synthesize a model of essential aspects of the ODE concept based on the empirical data. Our conceptual model comprise aspects of \emph{data value, data intrinsics, ODE governance}, and \emph{evolution}. Furthermore, we identify benefits and challenges related to the ODE concept, as well as possible actions to exploit the benefits and mitigate challenges in establishing ODEs\change{, and identify propositions for practice and further research}.

The rest of this paper is organized as follows. Related work is presented in Section~\ref{sec:RW}. The research method is outlined in Section~\ref{sec:method}. Section~\ref{sec:focusgroup} presents the focus group results and Section~\ref{sec:casesurvey} presents the case survey. In Section~\ref{sec:synthesis} we synthesize the results into a conceptual model and summarize practical implications of the findings. Threats to the validity are presented in Section~\ref{sec:validity} and the paper is concluded in Section~\ref{sec:conclusion}.


\section{Related work}\label{sec:RW}

Software ecosystems is a well established practice with supporting theoretical knowledge. Alves \emph{et al.} surveyed research on governance of software ecosystems and found 89 relevant papers~\cite{alves2017software}. They observed the importance of the platform owner and balancing of rights between owners and contributors. Our own research on open tools ecosystems~\cite{MunirIST2018} and product features in software ecosystems~\cite{LinakerJSS2018} focus on strategic choices on contributions, as a means for influence.

Oliveira \emph{et al.} identified four main organizational structures in their mapping study~\cite{Oliveira19}. In \textit{keystone-centric} ecosystems, actors are gathered around a central (keystone) actor who is the main provider of the shared data and orchestrates the ecosystem. In \emph{intermediary-based} ecosystems, a central actor is limited to intermediating data between data providers and data users. In \emph{platform-centric} ecosystems, actors interface each other mainly through a platform, commonly (but not limited to) data catalogs and portals. Finally, in \emph{market-place-based} ecosystems, there is a marketplace with rules and technical infrastructure that underpins the ecosystems. Comparing to software ecosystems~\cite{JANSEN20121495}, these structures commonly blend due to the importance of an underpinning technology platform and/or market. The platform ownership is highlighted by Dal Bianco \emph{et al.}~\cite{dal2014role} who distinguishes between \textit{keystone-centric}  and \textit{consortium-based} ecosystems. The former refers to cases where the governance resides at one single (keystone) organization, compared the latter where the governance is centered around an organization co-owned by the ecosystem's actors.

Attard \emph{et al.}~\cite{attard2015systematic} systematically surveyed literature on OGD and synthesized 75 papers with focus on governments as actors. The primary goal of government agencies is to increase transparency, although the access of information as such is brought forward as a benefit. However, the involvement by private companies or citizens as data providers is not addressed. Attard \emph{et al.} identified five categories of challenges for OGD, (1)~technical, (2)~policy/legal, (3)~economic/financial, (4)~organizational, and (5)~cultural, which seem to be relevant also for ODEs.

The potential innovation benefit from OGD ecosystems is discussed by Zuiderwijk \emph{et al.}~\cite{zuiderwijk2014innovation}. They advice how to create OGD ecosystems and define four key elements of an OGD ecosystem: (1)~government data provisioning, (2)~data access and licensing, (3)~data processing, and (4)~feedback to data providers. Furthermore, to get ecosystems into function, three additional elements are defined: (5)~usage examples, (6)~quality management system, and (7)~metadata. A survey among entrepreneurs indicate significant interest in OGD~\cite{lakomaa2013open}. However, the sustainability of funding is a threat to such entrepreneurial initiatives~\cite{KassenTransparencyNordic2017}. Case studies of OGD, e.g. by Dawes \emph{et al.}~\cite{dawes2016planning} and Styrin \emph{et al.}~\cite{Styrin2017}, indicate varying practices in different countries, and stress the socio-technical character of OGD ecosystems.

Accompanying ``The Value of Data'' report~\cite{CoyleValue2020}, the Bennett Institute for Public Policy also published a literature review~\cite{DiepeveenLiterature2020}. The review primarily takes an economical and policy view, but includes technical and governance implications. They identified literature in relation to (1)~categorization of data -- what is it and how can data be classified, (2)~benefits from data use, and (3)~barriers to data use. They further identified work on (4)~assessing the value of data, and particularly bring forward Mawer's unpublished piece on data valuation chains~\cite{Mawer2015}. Research on (5)~grades of openness are reported, ranging closed--shared--open data, which is further elaborated in their Data Spectrum model~\cite{CoyleValue2020}, and (6)~data trusts are in focus for literature on the need for intermediary organizations to handle data exchange (cf. intermediaries~\cite{susha2020towards}). (7)~The economic characteristics of data literature contests the idea that data is the new oil. Data can be used at a zero marginal cost (although data infrastructure and analytics do cost) and the same data can exist simultaneously in multiple places, which leads it to be considered as non-rivalrous in nature. As  consequence, companies tend to pile up their own data which becomes a barrier to data sharing. Finally, the literature review discusses (8)~public health data as an application domain for data use and value creation.

Enders \emph{et al.} surveyed the literature on the closed--shared--open data spectrum~\cite{Enders2019} to guide the selected revealing in open data. Using a qualitative content analysis of the literature, they derived six ``dataset metrics'' related to the data, which impact revealing decisions: (1) coreness, (2) currentness, (3) extent, (4) granularity, (5) interoperability, and (6) quality. Further they identified five ``decision criteria'' for data sharing: (1) competitiveness, (2) data misappropriation, (3) innovation opportunity, (4) legal, and (5) privacy. 

Susha \emph{et al.}~\cite{SushaBazaars2017} introduced the concept of Data collaboratives as {``cross-sector (and public-private) collaboration initiatives aimed at data collection, sharing, or processing for the purpose of addressing a societal challenge''}. The collaboratives highlight the need for data sharing and collaboration, yet is limited in scope to societal challenges, which may not always align with the interests of commercial actors in a data ecosystem. 

There are a few examples of ODEs with commercial actors. Anderson \emph{et al.}~\cite{anderson2019corporate} report on OpenStreetMap, a mature example of an ODE focused on map data, and how companies (referred to as corporate editors) contribute to the underpinning data project motivated by their different agendas, much like how OSS ecosystems function~\cite{LinakerEMSE2020}. The ecosystem is decentralized globally with overlapping communities on both national, regional and local levels~\cite{anderson2019corporate}. The copyright to the data is maintained by a UK-registered non-profit foundation which supports the community's activities\footnote{\url{https://wiki.osmfoundation.org/wiki/Licence/Community_Guidelines}}. The governance of the data is however managed separately among the individuals engaged in each community with a consensus-driven process for decisions-making and managing individual edits\footnote{\url{https://wiki.openstreetmap.org/wiki/Vandalism}}. 

Linåker \emph{et al.}~\cite{linaaker2020collaboration} and Rudmark \emph{et al.}~\cite{rudmark2019harnessing, RudmarkHICSS20} report on Trafiklab, an ODE with actors related to the public transport sector in Sweden. The underpinning platform provides real-time and static data on e.g., traffic and time-tables, which is collected from both public and private transport operators. The operators in turn own and manage the platform collectively through a co-owned company. Established in 2011, the ecosystem has a rich population of actors using and developing third-party applications based on the provided data. Particularly, Rudmark explores the need for open metadata standards for the exchange of data in an ecosystem~\cite{RudmarkHICSS20}.

In the literature, we did not find any research on the technical aspects of sharing data between corporations as a means to foster open innovation. This is confirmed by Oliveira \emph{et al.}~\cite{Oliveira19} who highlight the need for further research both with the areas of governance, management, and coordination of data ecosystems. Recently, however, Lis and Otto added to the knowledge through a case study on ecosystem governance \cite{Lis2020}, comparing intra- and inter-organizational aspects of data governance. Still, we see a need for more knowledge and guidance on this emerging concept.

\section{Research Method}\label{sec:method}
\begin{figure}[t]
    \centering
    \includegraphics[width=\columnwidth]{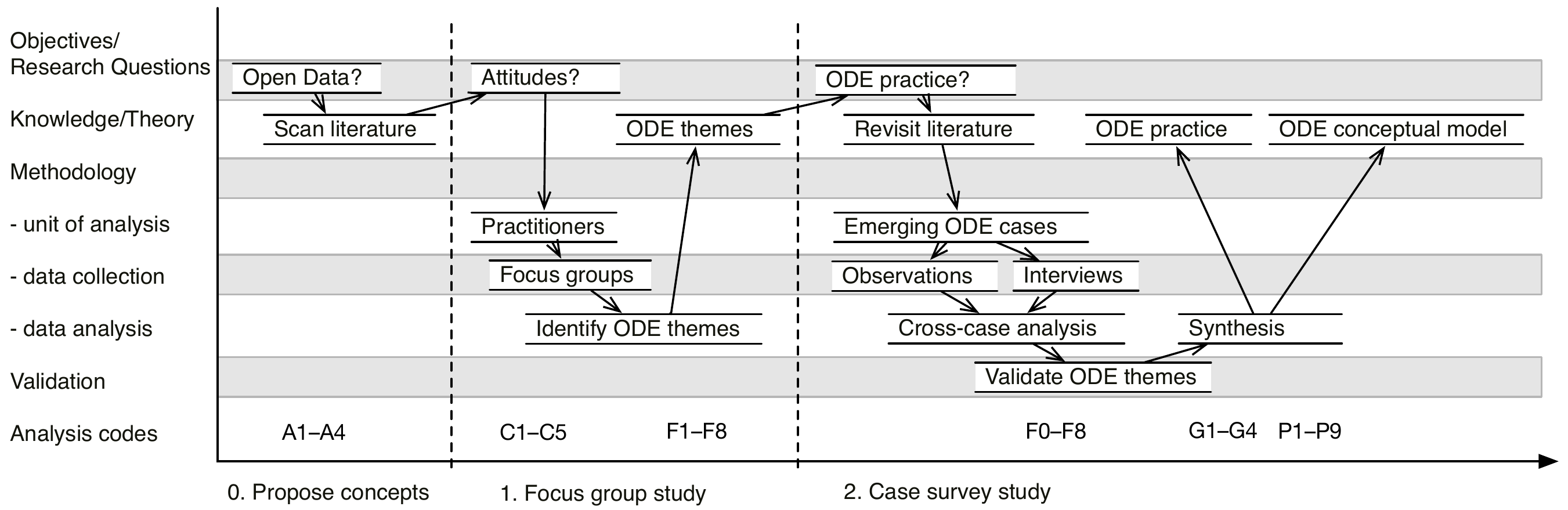}
    \caption{Overview of the research method. 0) The concept proposal phase corresponds to the ICSE NIER paper~\cite{Runeson19}, 1) the focus group study was first presented at the SEEA conference~\cite{RunesonSEEA2020}, while 2) this paper synthesizes both the focus group and the case survey studies.
    }
    \label{fig:method}
\end{figure}

We conducted a two-part qualitative study, as outlined in Figure~\ref{fig:method}, to explore the phenomenon of ODEs, using our previous conceptual proposal as a basis~\cite{Runeson19}. 

Our first step was to explore practitioners' views of themes related to collaboration around open data and their relevance to practice. We wanted to understand different organizations' attitudes to challenges and opportunities with ODE. We choose focus groups as our method for data collection~\cite{Kontio2008} and invited participants broadly from our network of commercial and public organizations to attend.

Our second step was to in-depth compare three emerging ODEs through a case survey~\cite{PetersenCaseSurveyChapter2020} with respect to the themes from the first step.
The case survey instrument was designed based on the outcomes from the focus group, to further study the ODE concept through observations and interviews in three cases and validate the initial themes. The data analysis integrates the findings from both the focus group and the case survey.

The research questions for our study are derived based on our earlier research on OSS and our hypotheses on the potential development  of ODEs~\cite{Runeson19}:

\begin{itemize}
    \item[RQ1] What data is produced and used within and shared among organizations?
    \item[RQ2] Which challenges and benefits can be expected within an ODE?
\item[RQ3] What actions are taken by emerging ODEs to address challenges and exploit opportunities?
\end{itemize}

Our research study is exploratory. We aim to understand the characteristics of the studied phenomenon rather than trying to assess the opinions of a certain population. We, thus, primarily employ qualitative research methods. 

\subsection{Focus Group Participants}

We invited participants to focus groups through our three primary professional networks; the regional ICT innovation cluster organization\footnote{\url{https://mobileheights.org}}, the university's collaboration network, and the research institute's corresponding contacts. Participants were invited through newsletters and e-mail lists, and could register for any of six workshop occasions in three different locations.

We organized three workshops in two locations based on the registration pattern -- in the end 27 participants from 22 organization, of which 14 companies, 7 public organizations, and 1 non-profit organization attended. We ran in total five focus groups of 5--8 participants (the first two workshops were divided into two focus groups) in March and April 2019. Below, we refer to them as FG1.1, FG1.2, FG2.1, FG2.2, and FG3.

The 15 private organizations operate in different domains (automotive, computer and chips, IoT, IT services, medtech and telecom), and was a mix of large, medium, and small enterprises. Even though participants are sampled by convenience, and thus not a representative sample in statistical sense, they represent a broad range of private and public organizations.

The participants had different roles in their organizations. Most participants had senior positions, for example, technical or middle management roles. From most organizations only one participant attended, although some large enterprises sent more than one participant. For further details on the focus groups, see our conference paper~\cite{RunesonSEEA2020}.

\subsection{Focus Group Data Collection}

We collected data during the workshops through note taking by the secretary, and having each focus group summarize their key findings as presentation slides.  
Each of the workshop sessions followed a similar scheme. We first broadly introduced the concept of ODE. We then split the participants into focus groups, which discussed topics related to ODE under three main questions:
\begin{itemize}
\item What type of data does your organization use or produce?
\item Which data can be shared? Under which conditions? With whom?
\item Which are the challenges and opportunities for sharing data? 
\end{itemize}

During the focus group sessions, we let the participants' scenarios for data drive the discussion as much as possible. In conjunction with the focus group sessions, the two groups reassembled, and a summary of each group was presented and discussed. The schedule for the focus group can be found in \ref{app:guide}.

\subsection{Case Survey}
We conducted a case survey~\cite{PetersenCaseSurveyChapter2020} of three cases of emerging ODEs. Case surveys are originally used as a secondary research method, integrating evidence from multiple, primarily published sources of evidence. In this study, we take the structured approach from case surveys of comparing a set of factors between a number of cases. However, the case survey is not informed by published case study reports, but by embedded researchers and interviews with case participants. 

The authors are embedded researchers in three ODE innovation projects related to industry 4.0, automotive, and the labor market, respectively. \change{We present an overview of the the cases in Table~\ref{tab:cases}, as well as the roles of the interviewees for each case.} The Swedish government invests heavily in support for innovation based on artificial intelligence and machine learning and such innovation projects are funded to catalyze triple helix collaboration between private, government and research organization, and these cases are examples of such. 

\begin{figure}
    \centering
    \includegraphics[clip, trim=.5cm 6.5cm 4cm 5cm, width=\columnwidth]{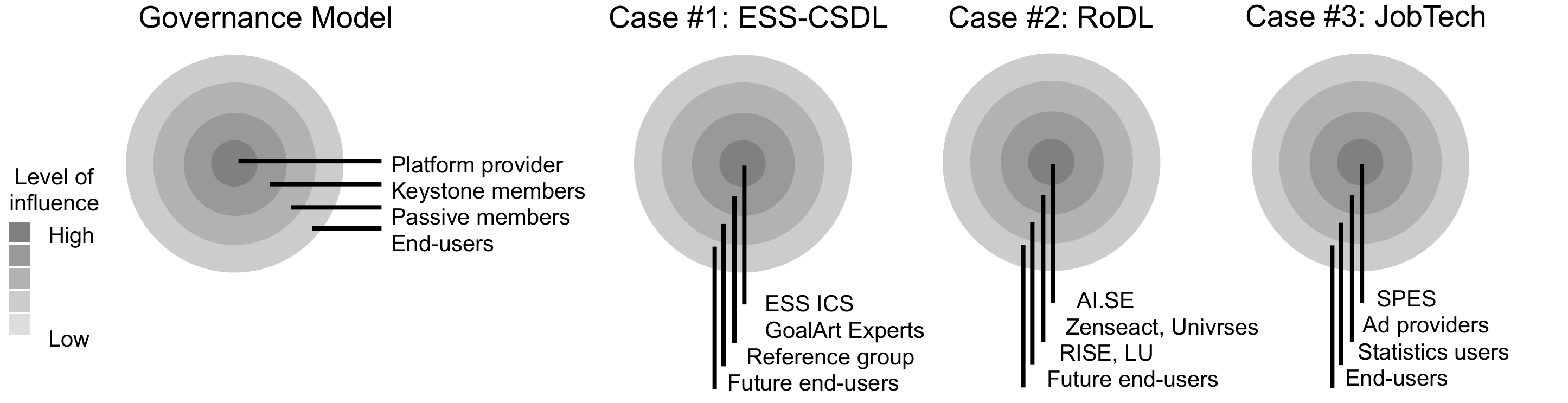}
    \caption{Overview of the governance model for the three surveyed cases, showing actors and roles according to Nakakoji \emph{et al}'s ``onion model''~\cite{nakakoji2002evolution}
    }
    \label{fig:OnionModel}
\end{figure}

We believe that the cases are relevant to further understand the ODE concept as they represent different domains, types of data, and type of actors. The three ODE's governance models in Figure~\ref{fig:OnionModel}, based on Nakakoji et al's ``onion model''~\cite{nakakoji2002evolution}. Each case is presented in detail below.

\begin{table}[t]
    \centering
    \footnotesize
    \caption{Summary of the three cases and empirical sources in the survey}
    \label{tab:cases}    
    \begin{tabular}
    {
       >{\raggedright\arraybackslash}m{2cm}
       >{\raggedright\arraybackslash}m{3cm}
       >{\raggedright\arraybackslash}m{3cm}
       >{\raggedright\arraybackslash}m{3cm}}
   \toprule
         &  \textbf{ESS--CSDL}   & \textbf{RoDL}  & \textbf{JobTech}\\
\midrule
Domain  &   Industry 4.0    & Automotive    & Labor market \\
\midrule
Type    &   Alarm data     &   Traffic video   & Job ads \\
\midrule
Researcher  &   First author    & Second author & Third author \\
\midrule
Interviewees    & Control systems specialist & Innovation project manager & Technical product manager\\
                & Alarm systems expert 1  & Deep learning engineer & Open data project manager\\
                & Alarm systems expert 2  &  & \\
 \bottomrule
    \end{tabular}

\end{table}

\subsubsection{Industry 4.0 data -- ESS--CSDL}
The European Spallation Source (ESS) is a European Research Infrastructure Consortium, currently constructing a multi-disciplinary research facility based on the world’s most powerful neutron source. ESS is a pan-European project with 13 European nations as members, including the host nations Sweden and Denmark, which in itself is a challenging collaboration endeavour. 

The ESS Control System Data Lab (ESS--CSDL) is focused on data from the Integrated Control System (ICS) which is designed to monitor and control approximately 100 000 devices in the ESS facility with a control system-related data flow of approximately 50 Gigabyte per second. ESS, although being a research facility, is on par with, or even at the top of, Industry 4.0 plants in terms of complexity and advancement. The ESS control system is currently under construction, but already today significant data is available from the gradually starting facility\footnote{\url{https://pos.esss.lu.se}}. The ESS--CSDL is scoped to establish an ODE for alarm data, aimed to catalyze innovation through data collaboration with external partners, particularly by sharing experiences, but also to make alarm data with high quality available for research and development in machine learning. 

In terms of Nakakoji et al's ``onion model''~\cite{nakakoji2002evolution} the core of the ESS--CSDL community is ESS ICS, complemented with alarms system experts from GoalArt\footnote{\url{http://www.goalart.com}}, see Figure~\ref{fig:OnionModel}, case \#1. Researchers from Lund University observe and advice the project with respect to data engineering and collaboration practices. The ESS--CSDL project has a reference group, composed of industry representatives from the domains of automation, robotics, chemical processing, packaging, and data science. There is also a member from a sister facility in Germany, DESY\footnote{\url{https://www.desy.de}} (Deutsches Elektronen-Synchrotron) established 1959, willing to share their experiences and learn back, and representatives from the university's innovation office. The reference group members constitute the Passive members in the ``onion model''.  The long term goal of the community is to advance Passive members to become Keystone members in order to serve End users, which are expected to be the customers and users of ESS and Keystone members, experiencing improved alarm systems in their Industry 4.0 plants.

\subsubsection{Road data -- RoDL}
The Road Data Lab (RoDL) is an innovation project which aims to establish an ODE where actors on and around roads can share data to enable machine learning studies, for example, sensor data from vehicles or road condition data collected from The Swedish Transport Administration. In the developments towards autonomous driving, the costs for collecting and annotating data for machine learning purposes has been identified as a significant challenge, and thus actors explore possibilities to reduce costs, still keeping their competitive advantage of the data.

The project explores technical, legal, and organizational issues in relation to sharing data among partners, as well as adding to its value. For example, one partner could provide the raw video data, and another partner could add annotations to the video data. 

The core of the RoDL community consists of AI Sweden (Swedish National Center for applied Artificial Intelligence\footnote{\url{https://www.ai.se/en}}) as platform provider, and Zenseact (an autonomous driving and safety software company, owned by Volvo Cars\footnote{\url{https://www.zenseact.com}}) and Univrses (a startup company with expertise on computer vision in smart city applications\footnote{\url{https://univrses.com}}) as keystone members, see Figure~\ref{fig:OnionModel}, case~\#2. Other project members include RISE Research Institutes of Sweden and Lund university, adding to the expertise in relation to open innovation, machine learning, and data ecosystems. There are ongoing discussions with organizations within the traffic/transportation domain to participate in the collaboration, e.g. The Swedish Transport Administration, who already runs a data sharing project with automotive vendors to collect friction data for safety purposes\footnote{\url{https://www.trafikverket.se/om-oss/nyheter/Nationellt/2019-02/moderna-bilar-hjalper-oss-att-halla-koll-pa-halkan/}}.

\subsubsection{Labor market data -- JobTech Dev Joblinks}
JobTech Development (JobTech) is a data ecosystem bringing actors together with the common vision of improving the digital match-making and guidance services on the Swedish labor market. The ecosystem, initiated in 2018, is facilitated by the Swedish Public Employment Service (SPES), see Figure~\ref{fig:OnionModel}, case \#3, who also develops and maintains the underpinning platform that consists of three main data sets: 
\begin{inparaenum}[1)]
    \item job ads related to the Swedish labor market, 
    \item resumés of individuals managed and controlled by the individuals themselves according to the MyData principles\footnote{\url{https://mydata.org}}, and 
    \item a taxonomy of skills and work titles, and the relationships in between.
\end{inparaenum}

The case examined in this study focuses on the job ads data set. Hence, when we refer to JobTech, we consider job ad data set and the actors connected to this specific data set. 

The ads are collected from ten large job ad providers related to the Swedish labor market. SPES collects the ads either through an open source scraping technology, or by calling the providers' API:s, when present. Once collected, the ads are transformed to a common format\footnote{\url{https://schema.org/JobPosting}},
 enriched with statistical identifiers for the specific job types announced,
 enriched with further meta-data to make to ad searchable,
 searched for duplicates which are marked and coupled,
 reduced to only contain job title and a link back to the original ad (''back-to-source''),
 and finally published on a public API maintained by SPES.

The reduction of ad data is the consequence of an agreement between the ad providers as they consider the ads as having a differentiating value. This way they still enable traffic back to their original ad, while gaining a wider reach as their ad is published through the common API.

The ad providers, whom can be considered as keystones in the ecosystem (see Figure~\ref{fig:OnionModel}, case \#3), range from small and specialized to large, international and general ad providers, for which the Swedish labor market only constitutes a small market. The goal, however, is to enable more providers to join after an initial demo phase (currently underway at the writing of this paper). There are also a number of actors, more focused on labor market statistics rather than match-making. These actors, including Statistics Sweden\footnote{\url{https://www.scb.se/en/}} and the analytics department of SPES, are organized in an advisory board, but contribute actively to improving and validating the enrichment process of the collected ads. Hence, they still have an explicit influence on the job-ad platform, although not as explicit as the ad providers as visualized in Figure~\ref{fig:OnionModel}.

\subsection{Case Survey Data Collection}
The codes from the analysis of the focus group study were turned into survey questions to be asked to each case, see \ref{app:survey}. Each co-author summarized answers for one case each and then interviewed 2--3 other project members (listed in Table~\ref{tab:cases}) to get additional perspectives on the cases. Notes from the interviews were collected in spreadsheets and then integrated to a free text survey answer for each question, which were iterated with the interviewees until they agreed on a correct response to each question. The data collection was performed in November 2020.  

The interviewees of the ESS--CSDL case are all involved in the innovation project. One interviewee has a PhD in engineering physics and now responsible for a ML project at ESS. The interviewee has been working with establishing ESS since 2009. The other two interviewees are experienced alarm specialists from GoalArt, with more than 20 years of experience from innovation and consulting on alarms in the process industry. They have PhDs in automatic control and electrical engineering, respectively, and are engaged part time in academia. 

The interviewees for the RoDL project were an innovation manager and a ML engineer from two different partners. The former is head of innovation collaboration with several years of external innovation projects. The latter is a PhD in computer science, now working with ML for computer vision applications. 

The interviewees from the JobTech project were the technical product manager, overall responsible for the technical development of the JobTech platform, and a project manager for multiple initiatives, including the job ad collaboration highlighted in this study.

\subsection{Analysis}

The qualitative analysis of the empirical data was conducted in three major phases, as indicated in Figure~\ref{fig:method} \change{where the last row shows the resulting codes for each phase}. The first phase analyzed the data from the focus groups, the second is a cross-case analysis of the survey data, and the third phase is a synthesizing analysis, including all data. 

In the first analysis phase, the findings from each focus group were briefly summarized after each workshop, structured according to the questions of the focus group.
After all the workshops, all notes were merged and then coded according to standard research practices~\cite{robson_real_2002} to identify \emph{ODE themes}. The coding and grouping of codes was performed by the second author. We started the analysis with \emph{a priori} codes, based on the topics for the focus group meeting (C1--C5). The coding was refined and synthesized in two iterations, resulting in eight final topic codes (F1--F8). After this process, conducted by the second author, the code structure was reviewed by the first author. Changes in the outcome were primarily about modification of terms and a more precise definition of the codes. \change{This is the analysis presented in our conference paper \cite{RunesonSEEA2020}.}

The second analysis phase focused on the material collected in the case survey. The first author performed a cross-case analysis~\cite{Cruzes14} for each of the codes, which had a dual goal. It ensured that the codes from the focus groups where valid for the new observations, and added richer data on the concept. The cross-case analysis was then reviewed by the co-authors. \change{This step resulted in one new code (F0) and the previous ones (F1--F8) being validated.}

Thirdly, the themes where synthesized into an \emph{ODE conceptual model}, including empirical findings and related literature. One top-down and one bottom-up synthesis model was created by the first and third authors, respectively, and then reviewed by the co-authors and integrated into one. The synthesis activity ended up in a conceptual model with four groups of ODE aspects (G1--G4). \change{Finally, for each of the ODE aspects, we synthesized our findings into propositions (P1--P9), which capture our initial understanding and outline directions for further research. }

The results from the first analysis phase, the focus group, is reported in Section~\ref{sec:focusgroup}. The results from the second analysis phase, the case survey, is presented in Section~\ref{sec:casesurvey} \change{together with the bottom-up synthesis model in Figure~\ref{fig:crosscase}. The synthesis is presented in Section~\ref{sec:synthesis} with an overview of how the codes emerged along the study in Figure~\ref{fig:concepts} and the resulting conceptual model in Figure~\ref{fig:conceptModel}.}


\section{Results from the Focus Groups}\label{sec:focusgroup}

\begin{table}[t]
\centering
\footnotesize
\caption{Final analysis codes}
\label{tab:codes}
\begin{tabular}{@{}lp{2.5cm}p{9cm}@{}}
\toprule
\textbf{ID} & \textbf{Code name} & \textbf{Code definition}\\
\midrule
F1 & Value of data & Potential business models, costs related to the collection and annotation of data, and business value of data\\
F2 & Value of collaboration & Conditions for and effects of collaboration around data\\
F3 & Data acquisition & Acquisition and brokerage of data\\
F4 & Relationships & Relationships between parties sharing data \\
F5 & Competition & Aspect of competition between parties sharing data \\
F6 & Quality & Quality of data, and what contributes to the quality \\
F7 & Maturity & How a data ecosystems may mature, with a particular focus on competence needs and standardization\\
F8 & Legal & Licensing and legislation for data\\ 
 \bottomrule
\end{tabular}
\end{table}

This section presents the results obtained through the qualitative analysis of the focus group sessions. We begin with types of data and then structure the rest of the section according to the topic categories, as defined by codes F1--F8 in Table~\ref{tab:codes}.

\subsection{Types of data}\label{sec:types}
In the focus groups several categories of data were identified, both based on the application domain that were represented and the characteristics of their data. We identified seven broad categories of types of data: Maps, Society, Position, Images, Sensors, Human, and Business, which we present below. Details about traces from data types to focus groups (FGx.y) can be found in our conference paper~\cite{RunesonSEEA2020}.

\emph{Map} data can be general, physical maps, with different layers of information. OpenStreetMap\footnote{\url{https://www.openstreetmap.org}} is an example of an ODE for maps. There are also companies, making business on map data (e.g. for military purposes) and there is a Swedish governmental authority, Lantmäteriet\footnote{Lantmäteriet maps the country, demarcates boundaries and helps guarantee secure ownership of Sweden’s real property. \url{https://www.lantmateriet.se/en/}}, that also partially is business based. Map data is often the backbone for various kinds of applications that have a connection to physical locations, for example transportation services.

\emph{Society} data includes all kinds of data related to the society. It may partly be seen as an extension to map data, adding information about buildings or technical infrastructure. Society data may also be dynamic, related to heat, electricity and different aspects of communication. However, it may also include information about events, regulations, decisions, as well as statistics on population or economical aspects of the society.

\emph{Position} data is related to maps, but is focused on the dynamics of transportation and individual movements. These can be seen as snapshots at a certain point in time, or time series for historical analyses and prediction.

\emph{Images} are data for training of machine learning applications. Faces and plants, were mentioned as examples for different machine learning applications. Image data may comprise individual captures or be sequences of images in a video stream.

\emph{Sensor} data refer to different kinds of measurement data from sensors, such as temperature, light, humidity in the environment, or sensors in a control system of a production plant. Sensors may be fixed or moving, the latter e.g. in a vehicle, where it may be connected to position data, which in turn may be connected to map data. 

\emph{Human} data is the most sensitive type, as it may be connected to many other types of data. Particular examples include behaviour, position, and health data. If human data is traceable to an individual or a small set of individuals, their privacy  may be threatened.  

\emph{Business} data may similarly be sensitive, as it includes customer data or is about the business as such. This data may also include usage data on the product/service provided, and thereby be connected to human data, e.g. the driver of a car.

In addition, the focus groups (especially FG2.1) discussed synthetic or generated data, as a source of data of different types for training of machine learning applications. These data are expected also to reduce the privacy issue, e.g. regarding training data for human face recognition. 

\subsection{Value of data}
The focus group participants expressed a sober attitude to the value of data, in contrast to evangelist statements on ``data as the new oil''. Participants clearly stressed that data have no value in itself, but must be connected to some business. One participant expressed that \textit{``big data means nothing if you do not have a business value''} [FG2.2]. Another participant expressed that it was difficult to put a number on the value of data [FG1.2]. This quite conservative position is particularly interesting as we invited participants to the workshop with focus on data -- i.e. the participants have a clear bias for an interest in data use and sharing. 

Several participants expressed that usage data from their customers and end users is important to improve their products and services [FG1.1]. All organizations in the workshops do collect data in one way or another. 

The opinions on `spillover' data differed, i.e. data which is not intentionally collected but gained as a by product of other data collection. Some argued for this data being well suited for sharing or selling, while another participant noted that the `gems' can be found in the part of the data that you did not intentionally collect [FG2.2]. It was noticed that \textit{``Google has a broad business model so they can cross-fertilize domains''} [FG3]. This was taken as an indication that Google's success is a kind of internal harvesting of spillover data.

\subsection{Value of collaboration around data}
When discussing business aspects of data, two types of costs were identified. Firstly, there are costs related to collecting data and ensuring its quality for the intended purpose. Data often need to be processed -- not seldom by humans -- to be useful. One such process concerns annotation of the data, which is key for machine learning. The participants see an opportunity to collaborate with other organizations in the annotation efforts in order to share the costs for the work. 

A second type of cost relates to data sharing, e.g., to ensure reliable and secure communication as well as additional mechanisms to filter out which data to actually share. Participants mentioned that \emph{``their systems are not prepared for sharing data -- neither with respect to APIs nor to content''} [FG2.2]. Furthermore, if the data is being shared as open data, additional resources are needed to validate and distribute the data. Hence, the participants agreed that collaborating in data ecosystems entails costs which needs to be matched by getting something in return. Collaboration without business value will not happen [FG3]. 

Collaboration around data an organization may also be a challenge. One of the municipalities participating in the workshop devoted it to political factors rather than technical ones [FG1.1]. Political factors include, structures, regulations, and ways of working become challenges for sharing data even within an organization. This was specifically raised from one of the municipalities in the study. They are, however, sharing ``master data'', i.e. information about inhabitants, addresses, and similarly. Contrasting this, there are certain legislation requiring municipalities to share data with Lantm\"ateriet and at the same time are required to pay for data from them as well [FG1.1]. In this case, the legislation is a challenge for the ODE. 

Participants pointed out that collaborating in ODEs might improve the quality of the data [FG1.2]. For example, if data shared with others is being annotated, this might add value. One participant pointed out, however, that not all data is equally interesting to collaborate around. They hypothesized that more general data is of greater value for collaboration, as opposed to very specific data [FG2.2]. Another participant mentioned that collaborating around data can be a way to increase market presence as well -- by getting insights and thereby the ability to build products and services for new customers [FG1.2]. Another potential opportunity is the pillar of open innovation -- by giving away some asset, the total market of the collaboration partners becomes bigger, a ``win-win situation''. This, however, requires adoption of open innovation principles. 

One participant stated that they might be more inclined to \textit{``trading the data with someone who is not a competitor''} [FG1.2]. Many participants reported having a concern that they give away a business value when collaborating on data. Therefore, they would rather collaborate with organizations that are not direct competitors. 

\subsection{Data acquisition}
Certain types of data can be purchased, such as market data and data collected by smart phones and apps. Marketplaces and data brokers exist, but even though participants had seen examples they were not particularly successful. \textit{``[E]specially in the insurance business and it was hard to get them fly"} [FG3]. However, even if a company wants to acquire data, e.g. annotated image data for machine learning, participants had observed a lack of available resources. 

Participants considered some type of data not possible and not likely to be available to buy. Often, companies are required to team up with others, perhaps even competitors, who are also collecting similar data to get access to more data. 

In the second workshop, participants speculated that there need to be public initiatives to build large data sets to support innovation [FG2.2]. The platform companies, such as Google and Facebook, have lots of data but it is under their control. Furthermore, for others to catch up on technology leaders in a certain domain, companies need to cooperate as the large platform companies have a head-start. 

\subsection{Relationships}
The participants pointed out that there must be a trustful relationship among the parties, in a data ecosystem. If an external party is responsible for the quality assurance of the data and the relationship is non-pecuniary, trust needs to be established by other means. Lastly, trust needs to be fostered and maintained. 

Participants in FG2.2 specifically mentioned that mutual sharing is key to create good relationships. That is, to be an ODE partner, you must give something away to receive something back. In FG3, participants also pointed out that there has to be a business rationale internally to motivate investments in sharing. 

Collaborating around data implies that data might be owned by other organizations. A participants stated that data that \emph{``owning your data you know it is correct''} [FG2.1] and thus you may be sure it is more reliable. This implies that the more important the data is for your business, the higher is the risk if the data is not owned by your own organization. 

\subsection{Competition}
Competition and competitors is a theme that recurred several times in the focus groups, which is natural as we discuss commercial aspects. One participant suggested that a way for smaller organizations to compete on a global market is to collaborate on data [FG2.2]. Otherwise, the large multi-national companies will have a too large advantage as they can collect and curate much more data. The participants suggest that forming local and regional clusters of collaborators may give an advantage. 

Another participant suggested that making data publicly available is another way of taking away the competitive advantage and, at the same time, contribute to the overall greater good for the society [FG2.2]. 

One hindrance for collaborating -- a participant in FG3 mentioned -- might be if the other organizations are better at turning the data into business value. Hence, it might be a disadvantage to collaborate on data or making it publicly available if other organizations are perceived as being faster. 

\subsection{Quality}

As data becomes more and more important for successful development and reliable operations, the requirements on data quality increase. Similar to ensuring the software quality, data quality also needs to be assured. Furthermore, just as reliable communication may be key for a system to operate as intended, data also needs to be reliable. 

Participants mentioned that having multiple sources of data can improve quality as well as sharing of costs related to the data [FG2.2]. Furthermore, if more companies are using the same data, inaccuracies are more likely to be discovered. Depending on the type of data, sometimes quality is about providing an exact fact -- e.g. a certain label -- while in other types of data, particularly measurement data, averaging over several sources gives more robust input. Quality criteria are hence different depending of the type of data.

Transparency also appeared as a topic in the second workshop [FG2.2]. To be able to trust data, it must be transparent how data were collected and curated, including any algorithms used in the processing. As opposed to OSS, it is not reasonable that the data is reviewed in its entirety. Rather, the procedures around the data should be checked. 

Even though we mostly have fast internet connection and cheap storage, it was brought up that the amount of data is growing very fast~[FG2.1]. Hence, there might be several practical challenges to sharing data as the amount of data grows. For some data, it might also be essential to have current data, which further aggravates this challenge of high speed communication. 

\subsection{Maturity}
Sharing software as OSS is an established practice, while ODE is in its infancy. In addition to the cultural resistance to open innovation as part of data ecosystems, participants mention that many of their systems are not technically prepared for sharing data [FG2.2]. Furthermore, they also state that even if sharing is technically possible, it is also required that the procedures for collecting and processing the data are standardized to ensure data is interpreted the same by different organizations. 

Participants in the second workshop pointed out that both the operational layer (those doing the actual work) and the strategic layer (those with power to decide) need to be aligned and understand data and sharing [FG2.2].

The lack of maturity was also brought up, in that several participants were missing suitable standards or APIs for data sharing [FG1.1 and FG2.1]. The municipalities, for example, mentioned that data is stored differently in different municipalities and the technical platforms and APIs also differ. Other organizations had similar observations. Furthermore, organizations are not used to sharing data with others. Hence, there are no processes or procedures for how to act in a collaborative setup [FG1.1], which is an organizational challenge. 

\subsection{Legal}
Legal aspects discussed in the focus groups were primarily related to GDPR\footnote{The General Data Protection Regulation (EU) 2016/679 is a regulation in EU law on data protection and privacy, which strengthens the right of the individual to its data. http://data.europa.eu/eli/reg/2016/679/oj} and uncertainties about how this regulation will be implemented [FG2.1 and FG2.2]. The uncertainty leads to a challenge, as collaborations might not happen when there is a reluctance to risk legal complications. 

There are also issues on license models for data. We have seen in the case of OSS that licensing is a complicated matter. Liability might also be impacted. If organization share data, depending on the license and the user agreement, liability might remain with the original data providing organization. 
Further, some participants perceive that legal uncertainties are more of a challenge than the technical ones [FG2.2]. Especially public organizations expressed more hesitance due to legal woes [FG2.2]. 

\section{Results from the Case Survey}\label{sec:casesurvey}
The discussions in the focus groups were based on experienced people forecasting what would happen when moving towards ODEs since the topic of ODEs is emerging rather than established. Thus, the case survey contributes to understanding the emerging practice of ODEs.

\begin{sidewaysfigure}
    \centering
    \includegraphics[clip, trim=0cm 1cm 0cm 1cm,width=\columnwidth]{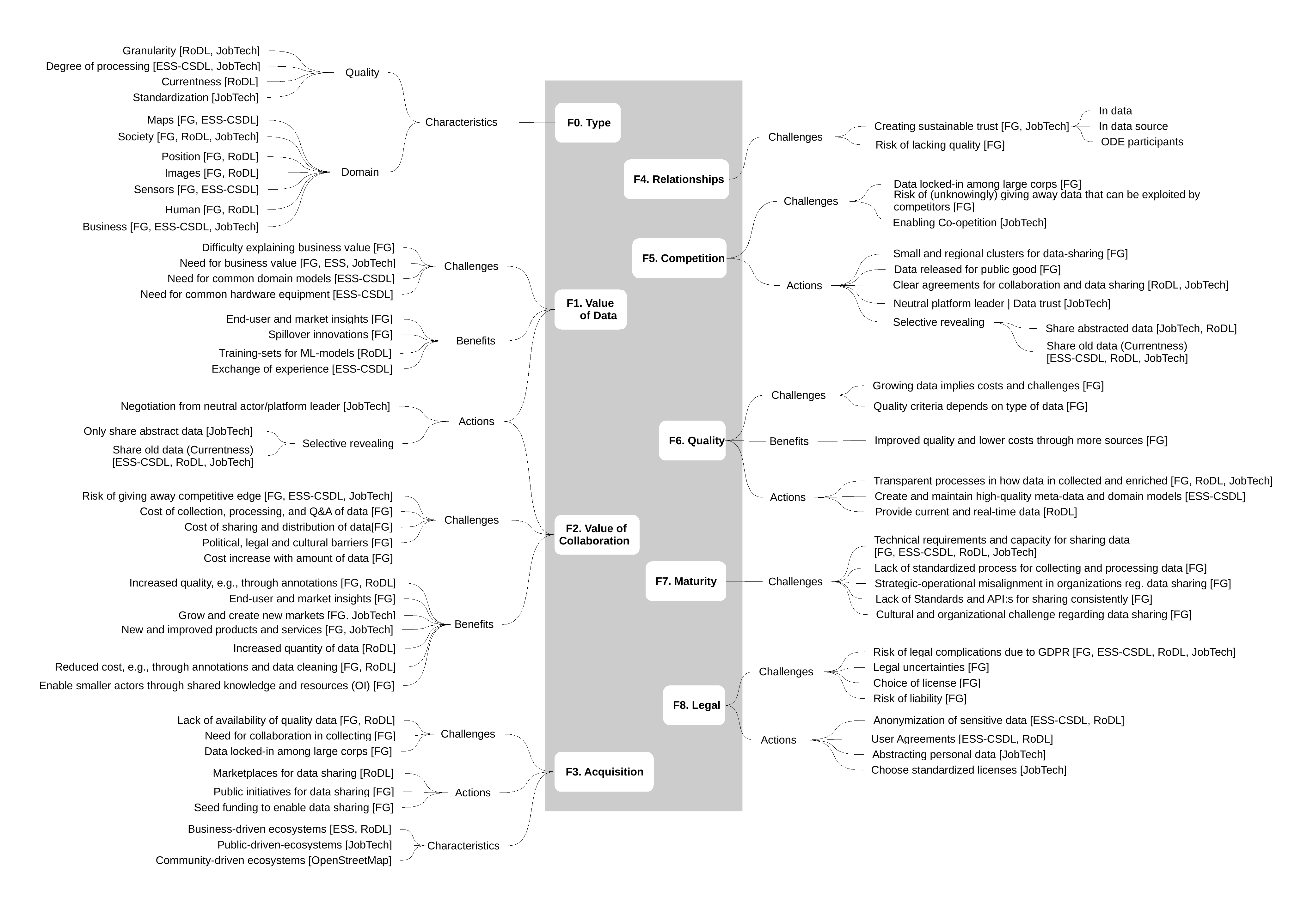}
    \caption{\change{Bottom-up analysis model, showing ODE aspects (F0--F8) identified in the focus groups [FG] and  the three surveyed cases [ESS--CSDL], [RoDL], and [JobTech], divided per sub-category (characteristics (RQ1), challenges (RQ2), benefits (RQ2), and actions (RQ3)).}
    }
    \label{fig:crosscase}
\end{sidewaysfigure}

Below we report the cross-case analysis of the eight topic categories (F1--F8) presented in Section~\ref{sec:focusgroup}, Table~\ref{tab:codes}. We validate the existence of the topic categories and \change{add \emph{types of data} as a separate category (F0) as it plays a central conceptual role. 
The bottom-up analysis model is presented in Figure~\ref{fig:crosscase}, where all observations related to the focus groups are marked [FG], and observations from the cases are marked [ESS--CSDL], [RoDL], and [JobTech], respectively. The observations are categorised according to general \textit{characteristics} (RQ1), \textit{challenges} and \textit{benefits} (RQ2), or \textit{actions} (RQ3)} in Figure~\ref{fig:crosscase}.

\subsection{Types of data}
\label{subsec:datatypes}
The three cases cover data related to all the identified categories, \emph{maps, society, position, images, sensors, human}, and \emph{business}. However, there is not a clear one-to-one mapping. The primary data in one case may fall into one category, while its meta-data or derived data may fall into another category, as shown below.

The ESS--CSDL data is primarily \emph{sensor} data, although at an aggregated level, where sensor information is processed into alarm signals. Furthermore, to make sense of the alarm data, a domain model is needed, which is a kind of logical \emph{map} over the facility. A time series of sensor data may, however, also constitute \emph{business} data, as it may indicate the health of the facility and thus also the efficiency of its operations. 

The RoDL data is primarily, \emph{images}, in the form of video streams connected to \emph{position} information collected from dedicated equipment either in dedicated vehicle or as an addition to standard vehicles.  
The images may also contain footage of \emph{humans}, as people are captured in the recorded videos and thereby connected to a point in space (\emph{position}) and time, via high precision GPS data. The images may also contain \emph{society} data, i.e. information about buildings and infrastructure, as well as information captured from road signs, that reflect e.g. parking rules or road conditions.

The JobTech data is primarily \emph{business} data, containing job ads from multiple sources. Furthermore, as job ads include names and information about contact persons, there is also \emph{human} data involved. Due to privacy reasons however, this data is removed in the enrichment process. Furthermore, as the complete pool of job ads provide information about market needs in different sectors of the job market, as well as a regional distribution over the country, it becomes \emph{society} data serving as foundation for labor market statistics.

The three cases validate the existence of all seven data types in the three cases. There is no case which directly focuses on \emph{map} data, but it is used as a domain or domain model for both RoDL (for its GPS coordinates) and JobTech (for its job locations). We also observe that the identified data types appear to be \emph{aspects} of the data, rather than orthogonal classifications, i.e., data may be of multiple types. Furthermore, the cases illustrate several other aspects of data, which may be orthogonal to the above, for example, \emph{granularity} (JobTech ads are shared at the job title and classification level only, and RoDL data may be shared with different granularity of annotations), \emph{degree of processing} (ESS alarms are processed sensor data, JobTech data is cleaned from duplicates), \emph{currentness} (RoDL data dates some time back), \emph{standardization} (JobTech adheres to the JobPosting standard\footnote{\url{https://schema.org/JobPosting}}) etc. 

\subsection{Value of data}
The business cases for the data may be in the data itself, but also to create value around the data or by extending the data volumes, and use it as an asset in collaboration around the data, see below in Section~\ref{subsec:collaboration}. However, there is a tendency to act as if the data on its own is valuable, and therefore particularly higher management and legal departments in the organizations are reluctant to sharing data.

ESS top management is committed to principles of open science, and the ODE draws on that principle to include operational data from the facility. In the processing industry, from which they have hired competency to build the facility, there is a strong tradition of keeping data closed. The expected benefits from ESS--CSDL is primarily related to sharing of experience. Sharing of data requires similar equipment between the partners, or at least standardized domain models, although some data for calibration has already been shared between ESS and its sister facility DESY.

The annotated videos constitute a business value in the RoDL case. The annotated videos are needed to develop the enabling technology, e.g., for autonomous driving, and the volume of data needed is huge. Thus, there is a potential business value in the annotated videos, which other actors may be willing to pay for. Hence, there is a value of the data, which can be increased by processing (i.e. annotating) which may be achieved in an ODE, i.e. adding value by sharing the data.

Publishing job ads and matching them to job seekers is the core business of several actors in the JobTech ecosystem. By publishing their abstracted ads with links back to the original ad, the can potentially increase traffic and exposure for the ad while maintaining the perceived differential value. Furthermore, the enriched ad data and connection to the standardized job type taxonomy enables them to improve ads, benchmark against competitors, and contribute to improved labor market statistics. Hence, there is a value in the sharing of data as such in the ODE which goes beyond the value of the data alone.

The case survey shows that the JobTech data is at the core of the business for some actors, and thus they only share abstract data. RoDL data is an enabler for the ODE members' business, constituting training data for their machine learning, although very costly. In the ESS--CSDL case, exchange of experience is considered more important than the data as such. The JobTech and RoDL cases confirm that there must be a business value in the data to motivate the collaboration, while in the ESS--CSDL case the primarily value is in the collaboration in itself. Cost for data annotation and classification is also a business aspect prevalent in the studies cases, particularly RoDL, while ESS--CSDL is expected to use unsupervised learning which does not need explicit annotations.

\subsection{Value of collaboration around data}
\label{subsec:collaboration}
All the three cases are examples of ODEs, where commercial actors are involved in sharing data more or less openly. ESS--CSDL and RoDL are innovation projects, with external funding to initiate the collaboration, while JobTech is initiated as a means for a governmental agency to fulfil its tasks, which includes interacting with competing commercial actors. 

The ESS--CSDL collaboration is still in its infancy, but the core partners as well as the passive commercial members (see Figure~\ref{fig:OnionModel}, Case \#1), find it relevant to keep pursuing the initiative, although so far only at the level of reference group members. In the view of becoming keystone members, the passive members foresee problems in relation to their customers and their own business and legal systems, rather than within the ecosystem. Their customers may be competitors, using e.g., the same robotics or automation systems in their production, and consequently reluctant to share potential business data with competitors. 

The RoDL collaboration envisions supporting innovation for autonomous driving and road safety. One motivation for the commercial partners for engaging in the ODE is to get access to more data than what they are collecting themselves. Also, if others are using their data, it is an indication that their data is relevant.

In JobTech, SPES has enabled a collaboration amongst the actors, whom to various degree consider themselves competitors, by sharing only abstract job data, forcing the end users to visit each partner's sites for the full information. This form of \textit{co-opetition}, i.e., collaboration between competitors, has evolved with time due to the efforts from SPES in finding a balance between perceived value and risk from the actors point of view. JobTech has also established APIs and a standardized taxonomy that helps collaboration.

The driving force for collaboration in the three cases is getting access to more data and knowledge, rather than the quality of the data as such. All the three cases have also spawned interest in collaboration beyond the specific data; ESS--CSDL sharing knowledge about alarm systems and industry 4.0 plants, RoDL about potential innovation in collaboration with other mobility actors, both for business and safety goals, and JobTech has spawn off a seminar series for other public agencies about working with open government data.

Interestingly, the collaboration around job ads turns \emph{business} data into \emph{society} data, as observed in Section~\ref{subsec:datatypes}. These job market data have traditionally been collected from regular surveys, and can now be observed in real time. In the ESS--CSDL case, the efforts to make data useful outside the organization is considered increasing the value of the data also within the organization, as it can be more easily accessed and interpreted by other than the data producers. 

In all cases, the \emph{currentness} of data is an important factor for the business aspects. Sharing old data is less sensitive than sharing real time data, and in some cases, data age rapidly, making it a matter of hours or days until they can be shared with little business risks.

The three cases demonstrate examples of collaboration around data, but it is still hard to assess the value of the collaboration. Whether or not they will be self funding beyond the innovation projects is a test of time for the concept. The cases also demonstrate that the ODE has to be balanced with competition between actors within and outside the ecosystems, which may involve lengthy negotiations.

\subsection{Data acquisition}
\label{sec:acquisition}
In neither of the cases, there are any established data brokers. In the RoDL case, there exist brokers trying to establish themselves, but there is a lack of suppliers of relevant data. Given the costs of collecting and annotating data for autonomous driving, the actors foresee a potential for a more established marketplace, but the domain has not reached such maturity yet.

This observation led us to extend the data acquisition topic to address generally how data is produced and by whom. All the three ODEs have public actors involved, although it is not about Open Government Data (OGD) in a strict sense. It is more of a Public--Private Partnership (PPP). Private actors produce data into the RoDL and JobTech ecosystems, public funding functions as a catalyst for the ESS--CSDL and RoDL cases, and the public agency SPES is at the core of the JobTech ecosystem. 

We further observed the distinction among the cases between \emph{public-driven, business-driven} and \emph{community-driven} data ecosystems. JobTech is an public-driven community, initiated and governed by SPES. RoDL and ESS--CSDL aim to be business-driven, although they now in their inception phase are supported by public funding. JobTech might also function independently of the SPES in the future, and thus also become a business-driven community. None of the three is community-driven, but we find crowd-sourced communities, like OpenStreetMap and WikiData, be examples of such ecosystems.

\subsection{Relationships}
RoDL and ESS--CSDL represent early stage ecosystems which are set up to explore the potential of ODE. These ecosystems do not stress the relationship between the actors and the culture is trustful and seen as joint exploration among the parties. 

The JobTech ecosystem is more mature, and involves competitors in the job ad market, and thus more challenging relations. All actors have a common interest in the Swedish labor market but with different perspectives. Ad providers with their business perspectives maintain a \emph{co-opetition} relationship with each other, while keeping a more open and collaborative relationship with SPES.
Originally, the ad providers were skeptical towards the collaboration but have easened with time. Generally the large and established ad providers have been more forthcoming to the collaboration than the smaller and younger ones.

\subsection{Competition}
The three cases represent different maturity of an ODE, and thus competition between actors is of different kinds.

The ESS--CSDL is characterized by the research culture of ``friendly competition'', influenced by open science principles. Collaboration exist with other research facilities world wide, and other large research facilities have demonstrated a willingness to share both experiences and specific data to support the development of ESS. However, it is not clear whether or not the funding agencies share the same view. The establishment of ESS included a fierce competition between up to five alternative locations in Europe. However, that was primarily on the political level, and now 13 countries collaborate on the facility.

RoDL is an innovation project, with two commercial partners involved. These partners are not direct competitors, but provide products and services in the same business domain. Still, it is important for the data providers that they consider potential competition in the future. Consequently, there is a large awareness of legal aspects and care taken to protect rights. 

The JobTech ODE involves competitors among the keystone members, and is based on an agreement to share abstract job ad data between them. However, the process to reach this stage took two years with one-to-one meetings between SPES and each actor, before they could have two competitors together. Actors have  expressed appreciation to have a neutral space facilitated by SPES as a neutral actor to discuss sensitive matters. Apparently, there are incentives to join the ODE, but it has to be well balanced with each member's business model.

The cases demonstrate clearly that competition is an issue in ODEs, and that proper agreements on the conditions for collaboration and data sharing must be in place. This is particularly true for the business-driven ecosystems, as defined in Section~\ref{sec:acquisition} above.

\subsection{Quality}
The data quality aspects were different in the three cases, but in neither of them data quality aspects was the driving force.

In ESS--CSDL, the quality standard of the core alarm data in the facility is very high. However, when sharing data, the meta-data and the domain model must be of high quality as well, to enable data interpretation outside its context. The data ecosystem may provide incentives to improve these data, which also improves the internal communication.

For RoDL, accuracy and consistency of annotations is important. The data ecosystem is expected to contribute to extended annotation efforts and thus improve or validate the data quality. Currentness of data may be a quality aspect for some applications (e.g. road condition monitoring) requiring recently collected or even real time data. This may conflict with business and privacy issues as discussed above in Section~\ref{subsec:collaboration}.

In the JobTech case, the quality is of lesser importance than quantity. The higher percentage of ads that are collected, the better overview can be created of the Swedish labor market for statistical purposes and to enable job-seekers to finds jobs they otherwise would not find. That said, quality aspects in terms of accuracy is still important, for example, in terms of assigning the right statistical identifier for the specific job type in each ad. This standardization makes the ads more precise and comparable across job ad providers. Statistics Sweden has contributed resources to validate the algorithms used in order for them to trust the data and use it as input for their labor market statistics. 

In all three cases, \emph{standardization} of data exchange formats is seen as an important quality aspect, as \emph{interoperability} is required for collaboration, but also has a positive effect internally within the organizations.  In ESS--CSDL, a domain model for the system and how its alarms are related is needed to make any use of the data. A major part of the initiation of the ODE was spent on compiling the meta-information about the data and the facility. In RoDL, domain models for objects in road scenes is at the core of exchanging and enhancing data within the ecosystem. The domain model may be a multiple levels of abstraction, like the data (e.g. a more fine grained model for internal use). In JobTech, the collected ads are transformed into the open and generally established JobPosting standard. Further, the ecosystem derives a taxonomy for job types and skills to enable data processing withing the ecosystem.

In contrast to standardization, \emph{transparency} is not seen as a major quality factor in relation to the external users of the ODEs, but internal transparency of sensors and calibration procedures (RoDL) and data collection and web scraping technologies (JobTech) are important to build trust in the data. The cases are not yet exchanging huge volumes of data, hence standardized API and storing technologies are sufficient, but they confirm the need of procedures and technical solutions to be in place.

\subsection{Maturity}
The cases confirm that ODE is a concept in its infancy. The most mature case (JobTech) was established in 2018, and the two others are still under establishment.

Among the three cases, there are experiences of working with OSS, and sharing data within organizations, but the external collaboration on data is new. The maturation process is visible in efforts towards standardization, as a basis for exchange at different levels of abstraction. In the less mature ODEs, basic procedures for data sharing (joint data storage with access control -- ESS--CSDL and RoDL) is sufficient, while for the more mature JobTech, exchange through APIs help manage both volumes and selected revealing of data. 

As we have earlier observed regarding OSS~\cite{MunirEMSE17}, data ecosystems seems to be dominantly initiated bottom-up. However, as it has an impact on both business and legal issues, management gradually get involved and accept or embrace the development.

\subsection{Legal}
\label{subsec:legal}
The cases confirm that legal issues are substantial in the establishment of a data ecosystem.
The data protection regulation (GDPR) is relevant for all three cases, although RoDL and ESS--CSDL do not intentionally collect personal data. Still, people are captured in road scenes and operator actions is an important aspect of the alarm meta-data. Different approaches to anonymization (e.g. blurring and other kinds of noise) as well as phrasing of user agreements, are used to mitigate these challenges.

In the JobTech case, sensitive personal information may be collected during the scraping process of ads. However, by abstracting the ad information in the data available through public APIs, no personal information on individuals representing the employer or union membership (which \emph{is} sensitive personal information according to GDPR) is published. However, it is still discussed internally at SPES whether there is regulatory support to collect the information in the first place, even though it is deleted within 6--12 hours after collection.

The above mentioned challenges on competition and other relationships between actors in a data ecosystem have to be regulated within a proper legal and contractual framework. While all cases would prefer standardized licences for data to reduce negotiation efforts, no established practice has evolved yet. Further, even with standardized open licenses, conflicts may occur as illustrated in more established ODEs~\cite{linaaker2020collaboration, rudmark2019harnessing} where map data cannot be transferred to OpenStreetMap (with Creative Commons license) as the ecosystem used the ODbL licence.

\section{Synthesis}\label{sec:synthesis}

\begin{figure}[t]
    \centering
    \includegraphics[width=\columnwidth]{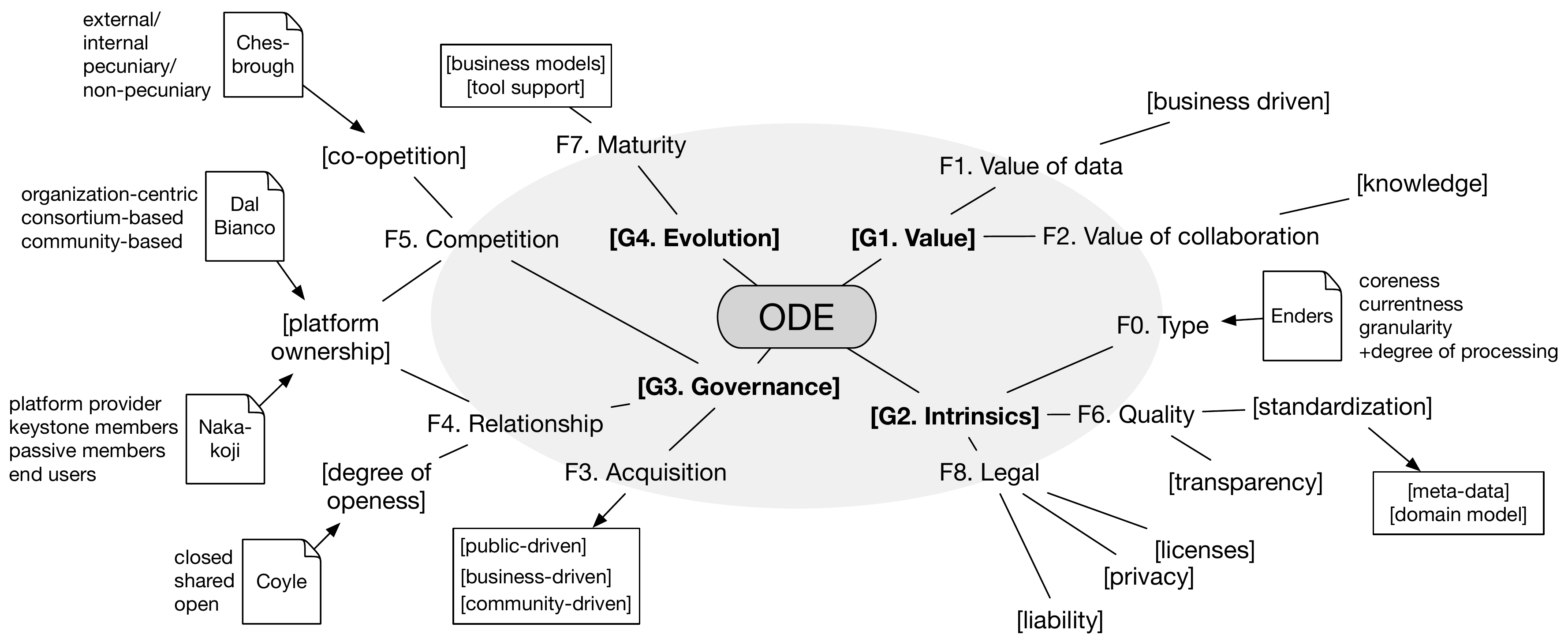}
    \caption{Conceptual model, depicting relations between the identified themes (F0--F8 and G1--G4), proposed aspects to consider (boxes), and core literature (boxes with cut corner). The inner, shaded part shows the relation between the aspects of the concept, while the outer part shows relations to findings and literature.}
    \label{fig:conceptModel}
\end{figure}

\begin{figure}[t]
    \centering
    \includegraphics[width=\columnwidth]{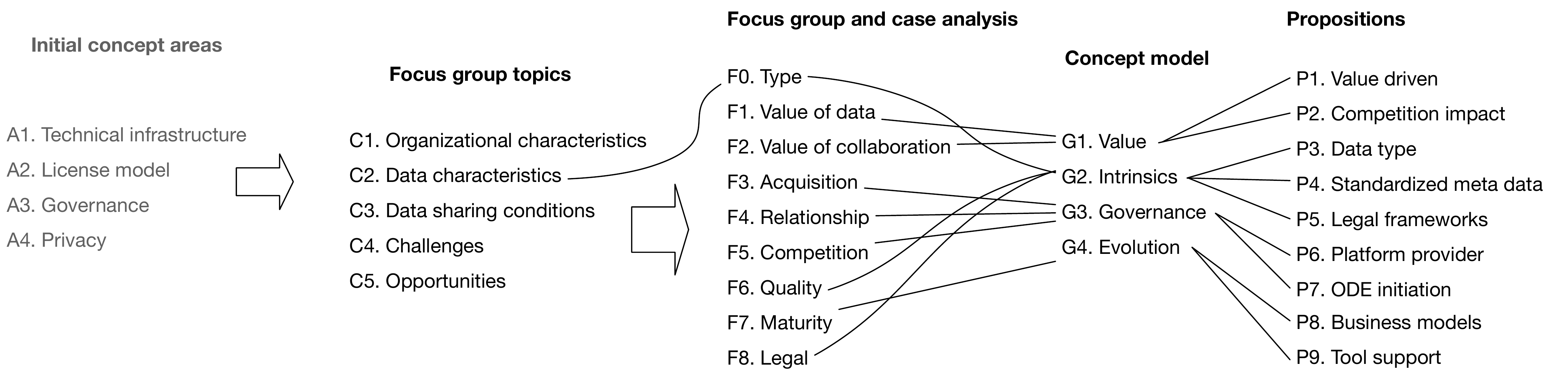}
    \caption{Overview of the evolution of the ODE concept throughout the analysis
    }
    \label{fig:concepts}
\end{figure}

Based on the analysis above, we further synthesize the findings into a conceptual model, which depicts initial relations between the identified themes as well as connections to some of the related literature. The  resulting conceptual model is presented in Figure~\ref{fig:conceptModel}, and the top-down synthesis, leading to its definition \change{and propositions for practice and further research}, is discussed below. \change{The emergence of the ODE concepts throughout the analysis is summarized in Figure~\ref{fig:concepts}.}

The eight codes (F1--F8, see Table~\ref{tab:codes}), emerging from the focus group analysis, and further explored through the case survey data together with the types of data (F0), were highly interrelated in clusters. Therefore, in our continued analysis we synthesized them into four higher level groups of aspects (G1--G4). 

\begin{enumerate}[G1]
    \item\textbf{Value.} The value of data (F1) and the value of collaboration around the data (F2) are two sides of the same coin. One or the other may be the primary value, but they are highly intertwined. 
    \item \textbf{Intrinsics.} Everything in an ODE relate to the data, by definition. However, some aspects are more related to the intrinsics, or internal characteristics of the data. Among those, we find the data type (F0) and data quality (F6). We also find legal aspects (F8) be tightly connected to data, although they also connect to governance of the ODE. 
    \item \textbf{Governance.} The relationship (F4) and competition (F5) are highly related, as they refer to different kinds of relations between actors. Further, the acquisition (F3) of data also depends a lot on relations between actors in the ecosystem and how they are governed. 
    \item \textbf{Evolution.} Finally, the matters of maturity (F7) are about the further evolution, since ODEs are in their infancy and in need for further research and development.
\end{enumerate}

Below, we discuss the four high level aspects, and the core findings and literature for each.

\subsection{Value}
The literature and the focus groups stress that the driving force for data ecosystems are the \emph{value of data} or the \emph{value of collaboration around data} for the actors. 
As two of the ecosystems in the case survey are still innovation projects, their long-term business goals cannot be assessed yet. However, the companies invest their own time in the projects, and thus at least have seen some potential business benefit, and they acknowledge that the collaboration around data is as important as the data itself -- at least in this stage. 

The case survey identifies different priorities in the business values in the collaboration. The ESS--CSDL puts the value of the collaboration before the pure exchange of data. They considered their own data being of sufficient quality and quantity for the primary purpose, while they seek experiences and knowledge about similar facilities and data. The RoDL partners see, at least short-term, that there is a value in getting access to more data by participating in the ODE. Long-term, the collaboration can be an end as well, not just a mean. The JobTech case, initiated by SPES, a government agency, has through negotiations managed to get multiple job ad providers on board, thus they have at least found the collaboration worth trying, even though we do not know details about their business considerations.

The concern of giving away business advantages was mentioned several times in the focus groups, i.e., creating more business value to others. For example, other organizations might be faster at developing their products and services or that other organizations might find business value in the data which you did not find. Therefore, they are more inclined to work with organizations which are not competitors. 

We synthesize our findings on value (G1) into two propositions:
\begin{itemize}
    \item[P1.] ODCs are driven by (F1) the value of data or (F2) value of collaboration.
    \item[P2.] The value of collaboration (F2) is impacted by the (F5) competition between actors.
\end{itemize}

\subsection{Intrinsics}

When classifying the data in the case survey, in relation to the seven types of data identified in the focus groups (\emph{maps, society, position, images, sensors, human}, and \emph{business}), we realize that they are aspects of data, rather than orthogonal classes. Consequently, data are not of a single type, but may have multiple aspects attached to data. The aspect may also change in the data processing (e.g., by combining job ads, \emph{business} data becomes \emph{society} data). Furthermore, the data may be collected with the purpose of being an \emph{image} at a \emph{position}, which also becomes \emph{human} data if a person happens to be in that position. This multi-type phenomenon implies that privacy and business issues may come into play, even though they were not originally intended.

We further observed in the case survey that other characteristics of data, may influence the decision to share or not. Some of these characteristics are already summarized by Enders \emph{et al.} as dataset metrics and decision criteria~\cite{Enders2019}. Depending on the \emph{granularity}, \emph{degree of processing}, and \emph{currentness}, the data may be shared or not with the ODE. Data, which abstracts away from individuals are less sensitive; processed data may involve less sensitive details, but may on the other hand also be more valuable, if combined with other data sources; old data may be less sensitive to share than current data, although data may age quite rapidly. There is also a technical aspect to currentness and granularity: large volumes of real time data require significant storage and transfer capacity, which may hinder sharing.

Lack of standards and technical infrastructure were often mentioned as a reason why data is not shared. Correspondingly, access to standardized meta-data and domain models within the ODE was seen as a significant quality attribute in all three surveyed cases. The model may be joint within the ecosystem, or be standardized in a wider sense.  This observation is well in line with Rudmark's work on open data standards~\cite{RudmarkHICSS20}.

\emph{Legal} issues are, as expected, manifold when establishing new types of data ecosystems. GDPR issues appear directly or indirectly in all cases, as humans are creating the data, and are mentioned or shown in the data, even when the human data is not wanted. Standardized licences are requested, but there are no established ones, except for the very open ODbL and CC licences, which are not free of problems as mentioned in Section~\ref{subsec:legal}.

\emph{Privacy} is a key concern when discussing data -- although peoples' practice still seem to be very relaxed towards sharing data through commercial platforms. However, seen from the perspective of a private company or a government authority, privacy issues seem to be taken very seriously. The JobTech case actively works on abstracting data to protect privacy.

Furthermore, \emph{liability} is unclear. What can be expected in terms of not only complying with license and privacy laws but also consequences if, e.g., incorrect data leads to problems when shared with others? We believe here are long-term policy questions to be addressed, as well as needs for creating an environment where it is accepted to take a legal risk and venture into uncharted territory. This is not found in the cases yet, although it is a part of the ongoing discussions in the RoDL case. 

We synthesize our findings on data intrinsics (G2) into the propositions:
\begin{itemize}
    \item[P3.] The type of data (F0) and its characteristics impacts the degree of openness (F4)~\cite{Enders2019}.
    \item[P4.] Standardized meta-data and domain models are core quality attributes for data~\cite{RudmarkHICSS20}.
    \item[P5.] Legal frameworks (F8) must be developed to support ODE evolution (G4).
\end{itemize}

\subsection{Governance}
The governance aspects of data ecosystems are more explored in the literature than the other aspects. Thus, we connect our observations to concepts already identified in the literature.
In Table~\ref{tbl:organizationalStructure}, we provide an overview of the three cases and, for comparison, also include OpenStreetMap (OSM) in the overview.

\begin{table}[t]
\caption{Characteristics of the data ecosystem's governance for the surveyed cases, with the addition of OpenStreetMap (OSM) for reference}
\footnotesize
\label{tbl:organizationalStructure}
\begin{tabular}{
>{\raggedright\arraybackslash}m{2cm}
>{\raggedright\arraybackslash}m{2.1cm}
>{\raggedright\arraybackslash}m{2cm}
>{\raggedright\arraybackslash}m{2.5cm}
>{\raggedright\arraybackslash}m{2.9cm}
}
\toprule
\textbf{ID} & \textbf{Initiation} & \textbf{Openness}& \textbf{Structure}  & \textbf{Data acquisition} \\ 
\midrule
JobTech & Public-driven & Shared, Open & Organization-centric & Platform provider, Keystone members \\
\midrule
ESS--CSDL &  Business-driven & Shared & Organization-centric & Platform provider \\
\midrule
RoDL & Business-driven  & Shared& Consortium-based & Keystone members \\
\midrule
OSM & Community-driven & Open & Community-based & All \\
\bottomrule
\end{tabular}
\end{table}

Regarding \emph{initiation} of data ecosystems, i.e. the driving force from the point of view of the platform, we found it be \emph{public-driven, business-driven} or \emph{community-driven} (see Section~\ref{sec:acquisition}). Even though the innovation projects in our case survey are publicly supported, they also have commercially funded parts, and the long term goal is to be commercially viable, i.e. being \emph{business-driven}. The \emph{public-driven} ecosystem (JobTech) is initiated since it is a task for the agency to boost the job market, but if the ecosystem gains enough commercial momentum, the agency may withdraw from it. None of the three cases is community-driven, and searching the literature for examples, we found only very large initiatives, like OpenStreetMap~\cite{anderson2019corporate}, which we here use as a point of reference. Maybe there are local or regional ones, that don't make it into the literature, but we still hypothesize that there must be a huge crowd to support a community-driven ecosystem to make it sustainable over time.

The relationships between actors in the surveyed cases range from an open science culture to competing business, while the ODEs still are in the mid of the spectrum of \emph{openness} (closed--shared--open~\cite{CoyleValue2020}). The data in all three studied cases is shared in a defined group of actors via authentication. The most developed case, JobTech, opens partial data to the public, for example job seekers. ESS--CSDL also has ongoing initiatives to share data openly to a broader group of actors, but whether or not it should be fully open or access-based is still under discussion. As a reference, OpenStreetMap has open access to anyone under the Open Database License (ODbL).  
The \emph{competition} concern within the ODE is similar as for any type of open innovation \cite{chesbrough2003open}, for example, open source software. While the core of ODE is open innovation, we believe that practices for collaborating around data is different than for OSS, and hence, need to be better understood to give organizations systematic approaches to evaluate this challenge.

Considering the \emph{structure} of the ecosystem, we differentiate between \emph{organization-centric}, \emph{consortium-based}, and \emph{community-based} ecosystems, as proposed by Dal Bianco \emph{et al.}~\cite{dal2014role} \footnote{To avoid confusion regarding the general use of the keystone concept in data ecosystems, we re-label  Dal Bianco \emph{et al.}'s \textit{keystone-centric} to \textit{organization-centric} ecosystems.} 
With \textit{organization-centric} ecosystems  we refer to cases where the governance resides at one single organization (as for ESS--CSDL and JobTech). With \textit{consortium-based} ecosystems, we consider ecosystems where the governance is centered around an organization co-owned by the ecosystem's actors (as for RoDL). OpenStreetMap, as presented in the literature, is a \textit{community-based} ecosystems where the governance is decentralized among the individuals who are members of the ecosystem.

Trusting data sources, and trusting other organization, to use shared data in a proper way is a concern for many of the focus group participants. Building trust in a distributed and multi-faceted network is hard, and thus some hypothesize that there can be a role for a centralized function to ensure the quality and reliability of data. For open government data, the government agency is the guarantee (as in the case of JobTech), while in peer to peer sharing we have seen few alternatives to the big tech players. \emph{Data trusts} -- ``a legal structure that provides independent stewardship of data'' --  are proposed as one solution by the Open Data Institute~\cite{CoyleValue2020}. 

The \emph{data acquisition} defines the actors supplying data to the ODE. We define the roles of the ODE actors in terms of Nakakijo \emph{et al}'s ``onion model''~\cite{nakakoji2002evolution}, as presented in Figure~\ref{fig:OnionModel} (platform provider, keystone members, passive members, and end users). We see that it can either originate from a platform provider as in ESS--CSDL, from the keystone members as in RoDL, or from both the platform provider and keystones as in JobTech. OpenStreetMap additionally acquires data from end users.

We synthesize our findings on governance (G3) into the propositions:
\begin{itemize}
    \item[P6.] There is a need for an independent platform provider to ensure trust in an ODE.
    \item[P7.] ODE initiation may be public-driven, business-driven, or community-driven.
\end{itemize}

\subsection{Evolution}
The concept of and strategies for ODEs are still in their infancy. Existing literature mostly address open data, as shared by public organizations -- Open Governmental Data (OGD)~\cite{attard2015systematic, zuiderwijk2014innovation}. The literature thus does not give support in defining strategies and processes for data ecosystems with commercial actors, which addresses a wider range of issues, such as business relationships and legal matters. A mapping study by Oliveira \emph{et al.}~\cite{Oliveira19} confirm the lack of industry focus among the data ecosystems research, indicating a knowledge gap regarding further evolution of ODEs with commercial actors.

Thus we have identified the need for better integration of ODEs into business models and management strategies, including cultural and organizational issues. Furthermore, tools supporting collaboration around and sharing of data is not standardized, in contrast to corresponding tools for software. Consequently, we have initiated further research in these areas\footnote{B2B Data Sharing for Industry 4.0 Machine Learning\\ https://portal.research.lu.se/portal/sv/projects/b2b-data-sharing-for-industry-40-machine-learning(4a81afc1-8ec1-4e1a-ab7e-4351e779820e).html}.

We synthesize our findings on evolution (G4) of ODEs:
\begin{itemize}
    \item[P8.] \change{It should be established }how to integrate ODEs into an organization's business model.
    \item[P9.] Tools to support ODEs and enable data sharing should be developed and standardized.
\end{itemize}

\subsection{Summary of Analysis}
In summary, our analysis of focus group and case survey data, integrated with existing literature on software and data ecosystem, has emerged into a conceptual model of ODEs, \change{presented in Figure~\ref{fig:conceptModel}. The evolution of the concepts along the study is summarized in Figure~\ref{fig:concepts}.} We propose the model be validated and extended in further research, and used by practitioners to guide their establishment of ODE practices.

\section{Threats to validity}\label{sec:validity}

Regarding external validity, focus group participants were selected using convenience sampling~\cite{robson_real_2002}, and thus statistical generalization is not an option. However, for this exploratory, qualitative study, the primary focus is on diversity, which we report in detail in our conference paper~\cite{RunesonSEEA2020}. Hence, our results are relevant although we cannot say which are more important than others. However, we might have overlooked some domain where ecosystems are more established, which we have tried to mitigate by reviewing related literature and reach out in broad, multi-domain industry networks. The same holds for the case survey study, where the cases are selected based on the authors' involvement in the cases. Still, the cases represent a variation over different domains and types of data, as reported in Table~\ref{tab:cases}, although we cannot claim any saturation over domains and data types, and thus propose more data ecosystems be studied.

A more significant threat is that we explore a topic that is still in its infancy, which is confirmed by a recent literature mapping study~\cite{Oliveira19}. Thus we collect opinions and hypotheses in the focus groups, rather than facts and experiences in relation to the ODE. Furthermore, as the constructs are not well defined, we might misinterpret the participants. In order to mitigate threats to construct validity, we gave a short introduction to the data  ecosystem concepts in the beginning of each workshop. Furthermore, among the focus group participants, significant experience with OSS was represented, which gave a frame for the open concept. Our main mitigation strategy is to launch the case survey, although two of the three studied cases are innovation projects, which also are a sign of infancy. However, as it is a new concept, the challenges of starting up an ODE are as relevant as the long term perspective.

On  internal validity, the coding of the focus group observations was performed by the second author who never had the secretary role. The first author had the secretary role for FG3. This procedure addresses researcher bias, as the codes are based on the notes by someone else and latter the codes are reviewed by the first author. Furthermore, we validated our preliminary results against the surveyed cases. This addresses confirmation bias, even though this is still a potential threat to the validity of our work. 

Threats to the internal validity in the case survey also relates to lack of independence, which is the flip side of the coin of being conducted by embedded researchers. Actions taken to mitigate the threat is to take two synthesis perspectives (top-down and bottom-up) and perform co-reviews of the cross-case analysis by the co-authors. Further, the data collection by the embedded researchers was made explicit through the case survey procedure of written responses to explicit survey questions (see \ref{app:survey}) which were scrutinized in member checking with other case members through the interviews.

\section{Conclusion and future work}\label{sec:conclusion}
We report the in-depth analysis of five focus group meetings and a survey of three cases regarding the concept of Open Data Ecosystems (ODE). Collecting input from 27 participants from 22 different private companies and public authorities, representing a variety of industrial and societal domains, provides a rich view of ODE factors. Furthermore, the case survey of three emerging ODEs gave a richer foundation and validated the findings from the focus groups. Based on our observations, we believe that ODE will be one way to realize open innovation, both in public--private partnerships, but also between multiple private actors. 

We conclude from our study that an ODE may use and produce many different types of data, although ODE is not yet a widespread practice (RQ1). Furthermore, we observe that data may be of multiple types at the same time, for example both \emph{sensor} data and \emph{human} data, one of which may be un-intentional. This leads to both technical and legal implications for the data ecosystems. Particularly, \emph{currentness}, \emph{degree of processing}, \emph{granularity} of data, and \emph{standardization} of meta-data and domain models are shown to be important aspects, which have an impact on the decision to share data or not.

Through our qualitative analysis of data, we defined a \emph{conceptual model} of  four main groups of aspects of an ODE (G1--G4). The \emph{value} (G1) lies in the data itself and in the collaboration around the data, which are the core business drivers for an ODE. The \emph{intrinsics}(G2) of the data embody its type, quality, and legal aspects, which influence the willingness and ability to open data. Further, standardized data and meta-data, particularly enable data exchange. ODE \emph{governance} (G3) may draw on research on open innovation in general, and software ecosystems in general. Finally, for the \emph{evolution} (G4) of ODEs, we identify business and tools support be essential areas.

A central factor, which is both challenge and benefits with ODEs (RQ2), is the value of the data in itself or in the collaboration around the data. OSS and software ecosystems is an open innovation practice which is successfully integrated into business, but it takes time to align it with legal and management operations. We observe that the introduction of ODE also takes time. Furthermore, human data is challenging from a legal point of view, and liability issues are also unclear. Trust in the data and the governance of an ODE is also raised as a challenge, both in the focus groups and in the literature.

In our analysis of emerging ODE practice (RQ3), we identified the ecosystem \emph{initiation} or driving force, its \emph{organizational structure}, the \emph{data acquisition}, and the degree of \emph{openness}, be differentiating governance characteristics. The surveyed cases of emerging ODE practice have all some kind of public involvement, although two of them they aim to be business-driven. The governance and data origin are the same keystone member of the ecosystem, which thus plays a central role. All three ODEs share data under constraints agreed within the ecosystem; only limited data is fully open. To build trust in the long-term development of an ODE, we see a need for a `neutral' role of matchmaker for the ecosystem. We also observe the demands on standardized domain models and data formats for data, to make sense across data ecosystem actors.

We synthesize initial patterns and need for more knowledge in nine \emph{propositions}. These call for further research, both on the practice broadly to theorize existing knowledge, but also specifically extend it on business and governance models, and standardized models and tools for data sharing.

We advice practitioners to start ODEs small with trusted partners, and establish technical and governance procedures, including standardized domain models. Anchoring the principles of open innovation in the organization is critical for the success of the ODE, as well as the overall business value of it. To enable co-opetition, it may be necessary to include a trusted party, e.g., a public entity that can facilitate initial negotiations and collaborations.

\section*{Acknowledgements}
We thank our collaborator Sofie Westerdahl of Mobile Heights for co-organizing the workshops. Thanks to the participants in the focus groups for their contributions. Thanks also to Dr. Markus Borg, RISE, and the anonymous reviewers of this journal for reviewing an earlier version of this paper.  This work was funded by the Swedish National Innovation Agency, VINNOVA, under grant 2018-04341 for groundbreaking ideas in industrial development, grant 2020-00025 for ESS Data Lab, grant 2019-05150 for Road Data Lab, and by the Swedish Public Employment Service.

\bibliographystyle{elsarticle-num-names}
\bibliography{references}

\appendix
\newpage 
\section{Focus group guide.}
\label{app:guide}
Below is a list of topics and questions to guide the workshops. They should not all be answered, rather it is kind of a checklist. For each of the three sections, spend 5 min individually on post-it notes, 10 minutes presentation in group, and 10 minutes discussion.
\subsection*{Individual notes}

What types of data does your company collect/handle/use? Examples?

\subsection*{Characteristics of data collection}
\begin{enumerate}
\item Are data collected as input to the development of the product/services or to the continued operation? 
\item What is the lead-time from a phenomenon occurs to that it can be observed in collected data? What is the lifetime of data? 
 \item 	How much effort needs to be put into processing the data before it is possible to make analysis? 
 \item To what extent is the analysis automatic? 
 \item 	To what extent are privacy issues related to the data collected?
 \item 	Are you using ML today? Big data? 
\end{enumerate}


\subsection*{Individual notes}
Which data can be shared? Under which conditions? To whom?


\subsection*{Sharing data}

\begin{enumerate}
\item Is the data shared with other organizations? 
\item Is the data a competitive advantage? Same domain/different domains?
\item Can it be a differentiator to share data -- and thereby being part of a community or ecosystem? 
\item What are the costs related to collecting data? 
\item How unique is the data to your organization? 
\item What would happen if you stop collecting data? 
\item Are you charging others to the data you are sharing to them? 
\item Do you make data publicly available without charge or other (direct) monetary incentives? Altruistic?
\end{enumerate}


\subsection*{Bridges and barriers}

\begin{enumerate}
\item 	Technical challenges in collecting data? Sharing data? 
--	Cloud, connectivity, bandwidth, security, etc. 
\item Legal barriers 
--	GDPR
\item Business
--	Competition, differentiation
\item Have you had security incidents where unauthorized individuals have gotten access to data? What type of data was accessed? 
\item Authenticity -- How do you ensure the data is authentic?
\item 	Costs -- 
Can cooperation reduce the cost of data collection? 
\end{enumerate}


\subsection*{Prepare for presentation}
Prepare summary in three slides according to sections above.

\section{Case Survey Instrument.}
\label{app:survey}

\subsection*{Types of data}
\begin{enumerate}
    \item Which types of data are used/shared/co-produced?
\end{enumerate}

\subsection*{Value of data}
\begin{enumerate}
    \item Which business value -- if any -- is there in the data?
    \item To what extent is data used to improve the actors's products and services?
    \item Is there any `spill-over' data involved?
    \item Is the cost data annotation a factor?
\end{enumerate}

\subsection*{Value of collaboration}
\begin{enumerate}
    \item What additional values are there in the collaboration?
    \item What political factors are at play?
    \item Are competitors an issue?
\end{enumerate}

\subsection*{Data acquisition}
\begin{enumerate}
    \item Are there any data brokers available?
    \item Are there any public data initiatives?
\end{enumerate}

\subsection*{Relationships}
\begin{enumerate}
    \item What characterizes the relationships within the ecosystem?
    \item Who owns the data?
\end{enumerate}

\subsection*{Competition}
\begin{enumerate}
    \item How does competition between actors influence the ecosystem?
\end{enumerate}

\subsection*{Quality}
\begin{enumerate}
    \item What quality attribute is important for the data?
    \item Does the ecosystem add to the quality of data?
    \item Does transparency play a role for trusting data? 
\end{enumerate}

\subsection*{Maturity}
\begin{enumerate}
    \item How mature are the organisations w.r.t. standardization of data, internally and for exchange?
    \item Are the strategic and operational layers aligned within the actors organizations?
\end{enumerate}

\subsection*{Legal}
\begin{enumerate}
    \item How does GDPR impact on the ecosystem?
    \item Which licenses are used?
    \item Are there any other legal issues?
\end{enumerate}

\subsection*{Other}
\begin{enumerate}
    \item Issues within the case, not covered by the above questions?
\end{enumerate}
\end{document}